\documentclass[%
    reprint, 
    superscriptaddress,
    amsmath,amssymb,
    aps,
    pra,
    floatfix,
]{revtex4-2}

\usepackage{graphicx}
\usepackage{xcolor}
\usepackage{siunitx}
\usepackage{bm}
\usepackage[colorlinks=true,linkcolor=blue,citecolor=blue]{hyperref}
\usepackage[capitalise]{cleveref}
\usepackage[mathlines]{lineno} 
\usepackage{mathtools}

\def\({\left(}
\def\){\right)}

\DeclareSIUnit\magnon{magnon}

\begin{document}
\title{High dynamic-range quantum sensing of magnons and their dynamics using a superconducting qubit}

\author{Sonia Rani}
\email{soniar2@illinois.edu}
\author{Xi Cao}
\author{Alejandro E.~Baptista}
\affiliation{Department of Physics, University of Illinois Urbana-Champaign, Urbana, IL 61801, USA}
\author{Axel Hoffmann}
\affiliation{Department of Materials Science and Engineering, University of Illinois Urbana-Champaign, Urbana, IL 61801, USA}
\affiliation{Materials Research Laboratory, University of Illinois Urbana-Champaign, Urbana, IL 61801, USA}
\author{Wolfgang Pfaff}
\email{wpfaff@illinois.edu}
\affiliation{Department of Physics, University of Illinois Urbana-Champaign, Urbana, IL 61801, USA}
\affiliation{Materials Research Laboratory, University of Illinois Urbana-Champaign, Urbana, IL 61801, USA}

\begin{abstract}
Magnons can endow quantum devices with new functionalities.
Assessing their potential requires precise characterization of magnon properties.
Here, we use a superconducting qubit to probe magnons in a ferrimagnet over a range of about 2000 excitations. 
Using qubit control and parametrically induced qubit-magnon interactions we demonstrate few-excitation sensitive detection of magnons and are able to accurately resolve their decay.
These results introduce quantum circuits as high-dynamic range probes for magnons and provide an avenue toward sensitive detection of nontrivial magnon dynamics.
\end{abstract}
\maketitle


In hybrid quantum magnonic circuits, superconducting qubits are coupled to collective spin excitations in magnetic materials \cite{tabuchiQuantumMagnonicsMagnon2016,yuanQuantumMagnonicsWhen2022}.
Large spin densities enable strong coupling between collective spin modes and qubits, allowing the detection and manipulation of single magnons~\cite{tabuchiCoherentCouplingFerromagnetic2015,lachance-quirionResolvingQuantaCollective2017,lachance-quirionEntanglementbasedSingleshotDetection2020,xuQuantumControlSingle2023}.
Introducing magnetic degrees of freedom in this way can endow quantum circuits with new functionalities, such as nonreciprocity~\cite{wangLowLossFerriteCirculator2021,owensChiralCavityQuantum2022,wangDispersiveNonreciprocityQubit2024c, zhangBroadbandNonreciprocityEnabled2020, kimMagneticfieldControlledSwitchable2024,hanBoundChiralMagnonic2023} or frequency transduction~\cite{hisatomiBidirectionalConversionMicrowave2016, osadaCavityOptomagnonicsSpinOrbit2016, zhangOptomagnonicWhisperingGallery2016}.
The realization and utility of such applications hinges on an understanding of the limitations imposed by magnon integration, and thus requires the ability to precisely characterize magnon properties.

A promising approach for sensitive magnon characterization is to use a superconducting qubit as a quantum sensor~\cite{wolskiDissipationBasedQuantumSensing2020},
and it is an intriguing question which magnon properties can be detected effectively in this way.
Experiments to date have focused on the single or sub-single magnon regime.
There is, however, a considerable interest in investigating magnon dynamics that deviate from those of the harmonic oscillator and that could be observed at larger magnon numbers.
For example, damping mechanisms beyond the phenomenological Gilbert description~\cite{bajpai_timeretarded_2019,lomonosov_anatomy_2021,unikandanunni_inertial_2022} or including quantum effects~\cite{yuanQuantumMagnonicsWhen2022,yuan_pure_2022} should be considered.
In addition, nonlinearities manifest at large occupation numbers~\cite{zheng_tutorial_2023,wang_bistability_2018,lee_nonlinear_2023}.
For a quantum sensor to elucidate such effects it would be important that it can detect an extended range of excitations and their dynamics.
We are therefore interested in the question to what extent a superconducting qubit can act as a quantum probe of large magnon numbers, and accurately and sensitively resolve occupation and decay.

Here, we demonstrate quantum sensing using a superconducting transmon qubit that is weakly coupled to a ferrimagnetic sample via a microwave cavity. 
Exploiting the dispersive frequency shift and shot-noise induced dephasing of the qubit we have realized magnon-detection of up to $\sim$ 2000 excitations at a sensitivity level of a few $\SI{}{\magnon\per\sqrt{\hertz}}$. 
We then demonstrate two complementary approaches to resolve magnon dynamics.
Combining high-sensitivity magnon counting with fast qubit control we have measured the time-dependence of the dispersive shift, allowing us to resolve magnon decay accurately and with high sensitivity.
In addition, we have used parametric pumping to engineer a resonant interaction that results in magnons acting as controllable bath for the qubit; in this way, noise properties of the magnon mode are mapped onto qubit relaxation. 
This approach allows infering the magnon decay rate as well as the steady-state population, and has the potential to resolve decay rates that exceed qubit control speeds.
These results demonstrate that superconducting qubits can be used as sensitive and high-dynamic range quantum probes for magnetic excitations and their dynamics. 

\begin{figure}[t]
\centering
\includegraphics{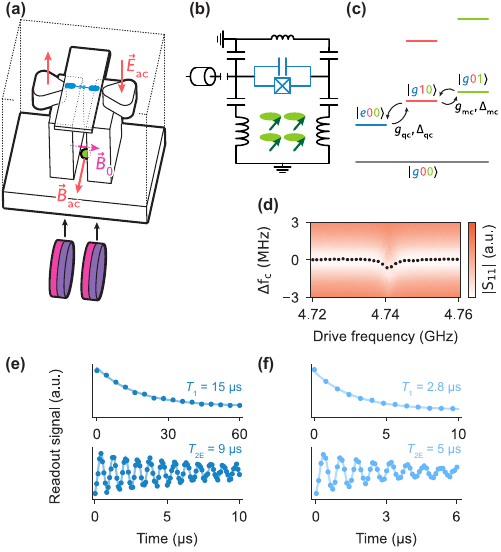}
	\caption{
        \textbf{Experiment overview.}
        (a) Schematic of the cavity.
        Transmon qubit (blue) couples to electric field of the cavity mode, YIG (green) to the magnetic field.
        Magnets (purple) generate a static field $\vec{B}_0$.
        (b) Equivalent circuit of the system.
        (c) Energy level schematic. 
        Qubit (blue) and cavity (orange) are coupled with strength $g_\text{qc}$ and are detuned by $\Delta_\text{qc}$. 
        Cavity and magnon mode (green) are coupled with strength $g_\text{mc}$ and are detuned by $\Delta_\text{mc}$.
        (d) Cavity spectroscopy vs.\ frequency of a drive applied near the magnon resonance frequency.
        Black dots: cavity frequency obtained from fits.
        (e) Qubit $T_1$ (top) and $T_\text{2E}$ (bottom) without magnets, and
        (f) with magnets inserted.
    }
\label{fig:fig1}
\end{figure}

Our experiment design is centered around a lumped-element cavity (\Cref{fig:fig1}(a) and (b)), motivated by the following considerations.
Quantum magnonic devices inherently face the challenge of combining superconducting circuits and magnetic fields.
Magnon modes in the few-GHz frequency range require a magnetic field bias; externally applied magnetic fields, however, suppress superconductivity and cause decoherence.
While field-compatible high-Q superconducting resonators have been coupled to magnons \cite{huebl_high_2013,li_strong_2019, hou_strong_2019, xiongCombinatorialSplitringSpiral2024}, contemporary superconducting qubits are still highly sensitive to field alignment \cite{schneiderTransmonQubitMagnetic2019, krauseMagneticFieldResilience2022}.
Past quantum magnonics experiments have overcome this challenge by combining externally applied magnetic fields with sophisticated partial shielding solutions \cite{tabuchiCoherentCouplingFerromagnetic2015, xuQuantumControlSingle2023, wangLowLossFerriteCirculator2021}.

A promising alternative is to employ local magnetic fields \cite{owensChiralCavityQuantum2022}.
We have designed a bus cavity that permits placing a qubit in a zone of near-zero field to retain high coherence and mediates coupling to a locally biased magnetic sample.
The electric and magnetic fields of the differential mode of this cavity are spatially separated \cite{angerer_collective_2016a}, making it straightforward to couple the transmon qubit capacitively, and the magnetic sample inductively to the mode.
We have used a ferrimagnetic yttrium-iron-garnett (YIG) sphere as an exemplary magnetic system to test our sensing approach, and focus our efforts on its Kittel mode \cite{cherepanovSagaYIGSpectra1993}.
The sphere is located near the maximum oscillating magnetic field;
magnets are placed in close vicinity, providing a local magnetic field bias (\cref{fig:fig1}(a)).
Because a static field is sufficient for our purposes, we have chosen neodymium permanent magnets producing an estimated magnetic field of about $B_0 \sim \qty{172}{\milli\tesla}$ at the sphere.
A fixed-frequency transmon qubit is located on a chip that is placed near the differential electric field, thus coupling capacitively to the cavity.
The anticipated magnetic field strength at the qubit location is $< \qty{5}{\milli\tesla}$ 
(Supplementary Materials).
Our sensing approach is based on a weak dispersive coupling between qubit and magnon mode, realized with mode frequencies and couplings as illustrated in \cref{fig:fig1}(c).
The transmon is coupled to the cavity in the usual dispersive regime \cite{blaisCircuitQuantumElectrodynamics2021e}, with coupling strength much smaller than mode detuning, $g_\text{qc} \ll \Delta_\text{qc}$.
The cavity-magnon coupling, $g_\text{mc}$, and detuning, $\Delta_\text{mc}$, are placed in a similar regime.
The fourth-order nonlinearity of the transmon qubit then results in a weak dispersive interaction of the magnon mode with both the qubit and the cavity.
The interaction is characterized by $\chi_\text{qm}$ ($\chi_\text{mc}$), the frequency shift of the qubit (cavity) per magnon excitation occupying the Kittel mode 
(see Supplement for details on the model).

We have first verified couplings and coherence properties at a temperature of $\qty{10}{\milli\kelvin}$.
To determine the resonant frequency of the magnon mode, we exploit the cavity dispersive shift, $\chi_\text{mc}$.
We applied a drive near the expected magnon mode resonant frequency, $\omega_\text{m}$. 
At resonance the drive populates the mode and shifts the cavity frequency (\cref{fig:fig1}(d)), establishing $\omega_\text{m}$ 
and confirming the nonlinear couplings introduced by the transmon.
We have further measured qubit relaxation and (Hahn echo) coherence times (\cref{fig:fig1}(e)).
Qubit coherence is reduced compared to measurements without magnets (\cref{fig:fig1}(f)), which we attribute to residual stray fields that can be overcome by optimizing geometry 
(see Supplement for details on stray fields).


\begin{figure}[t]
\centering
\includegraphics[]{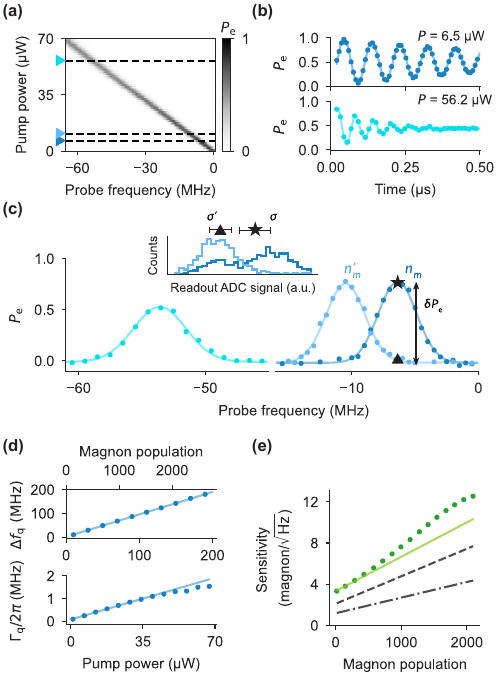} 
	\caption{
    \textbf{Magnon counting.}
    (a) Qubit spectroscopy as a function of pump power applied to the magnon mode.
    Probe frequency is relative to $\omega_\text{q}/2\pi$.
    (b) Ramsey decay of the qubit. A pump is applied on the magnon mode during the qubit evolution time.
    (c) Qubit spectrum for different pump powers (colors correspond to markers in (a)).
    Solid lines are Gaussian fits.
    $\delta P_\text{e}$ is the difference in qubit population after a $\pi$-pulse at $\sim \omega_\text{q}/2\pi -\qty{7}{\mega\hertz}$ with two magnon numbers prepared. 
    Inset: Corresponding qubit readout histograms. 
    (d) Stark-shifts and dephasing rates as function of pump power.
    Lines are linear fits.
    Dephasing at large powers deviates from linear as discussed in \cite{gambettaQubitphotonInteractionsCavity2006}.
    (e) Sensitivity as function of population (green dots).
    Green solid line: simulation.
    Difference between data and simulation arises from control imperfections at Stark-shifted qubit frequencies.
    Gray dashed line: simulation of ideal qubit case (no intrinsic decoherence).
    Gray dash-dotted line: simulation of ideal qubit and fastest-possible duty cycle.
    }
\label{fig:fig2}
\end{figure}

We have then characterized the qubit as a detector of the average Kittel mode occupation number, $\langle n_\text{m} \rangle$.
Due to the dispersive interaction, we can `count' magnons by detecting frequency shift and dephasing of the qubit.
As shown in \cref{fig:fig2}(a) and (b), as we increase the steady-state magnon population via a resonant pump tone, the spectroscopic response of the qubit shifts from $\omega_\text{q}$ to lower frequencies and the qubit dephases more rapidly.
To quantify the qubit's performance as sensor, we are interested in the sensitivity $S$ with which we can resolve $n_\text{m}$ magnons.
Sensitivity is defined such that, for a given $S(n_\text{m})$ and with unit signal-to-noise ratio (SNR) in one second of total measurement time, we can discriminate at best between $n_\text{m}$ and $n_\text{m}' = n_\text{m} + S(n_\text{m})$ excitations \cite{degenQuantumSensing2017}.
We obtain the signal as follows (\cref{fig:fig2}(c)).
We first prepare either $n_\text{m}$ or $n_\text{m}'$ magnons, and then apply a $\pi$-pulse at frequency $\omega_\text{q} + n_\text{m} \chi_\text{qm}$, the Stark-shifted qubit frequency for $n_\text{m}$ magnons. 
The signal is the difference in qubit excited-state probabilities between the two cases, $\delta P_\text{e} = P_\text{e} - P_\text{e}'$.
The noise arises from the uncertainty in determining $P_\text{e}$ with a given number of samples.

Quantifying sensitivity requires a calibration of magnon number as function of pump power, as well as the dispersive shift per magnon, $\chi_\text{qm}$. 
Both quantities can be obtained independently by measuring the qubit Stark shift and dephasing rate as a function of the magnon drive amplitude.
The shift in measured qubit frequency is given by $\Delta f_\text{q} = (\chi_\text{qc}/2\pi)n_\text{m}$. 
Additionally, fluctuations in the magnon number dephase the qubit at a rate $\gamma_2^\text{m}$ that is proportional to $n_\text{m}$~\cite{gambettaQubitphotonInteractionsCavity2006}.
We have obtained these quantities from qubit spectroscopy and Ramsey decay, yielding a calibration for dispersive shift and magnon occupation (\cref{fig:fig2}(d)) 
(See 
Supplement 
for details on the calibration).

Having established the magnon occupation, we determine the sensitivity as function of magnon number, $S(n_\text{m})$.
For each pump power applied, we fit the qubit spectroscopy signal and interpolate over drive power to obtain the sensing signal for arbitrary $n_\text{m}$ and $n_\text{m}'$.
Similarly, we estimate the noise for discriminating between any $n_\text{m}$ and $n_\text{m}'$;
the noise arises from the uncertainty in qubit readout, characterized by standard deviations $\sigma$ and $\sigma'$ (\cref{fig:fig2}(c)).
Together with the duration of a single execution of the experiment sequence, we obtain the conditions for unit SNR in one second, and thus the sensitivity (see 
Supplement).
The resulting $S(n_\text{m})$ from this procedure is plotted in \Cref{fig:fig2}(e), demonstrating a resolution of a few $\SI{}{\magnon\per\sqrt\hertz}$ over a $\sim$ 2000~magnon dynamic range; we set this maximum based on the finite control bandwidth of the electronics used in the experiment.

Limitations on sensitivity and dynamic range can be understood as follows. 
The dispersive coupling $\chi_\text{qm}$ is the central design parameter that determines sensitivity; it is the rate at which the qubit can acquire information about the magnon population.
Our choice of a small $\chi_\text{qm}$ is a central difference to earlier work in which a large coupling was used to resolve populations $\ll 1$ \cite{wolskiDissipationBasedQuantumSensing2020}.
In contrast, larger magnon numbers can be detected by a smaller $\chi_\text{qm}$.
As the magnon population increases the qubit dephases more rapidly, resulting in a broader qubit linewidth and a signal that is more difficult to resolved.
Range is thus limited by the technical challenge in measuring the shifted and broadened qubit spectrum with sufficient contrast.
A smaller $\chi_\text{qm}$ could thus always be used to further increase range at the cost of sensitivity.
Improvements to sensitivity are possible through qubit readout or faster measurement cycles, for example using active qubit reset \cite{geerlingsDemonstratingDrivenReset2013}.
Projected sensitivity limits due to improved qubit coherence and duty cycle are shown in \cref{fig:fig2}(e).


\begin{figure}[t]
\centering
\includegraphics[width=3.34in]{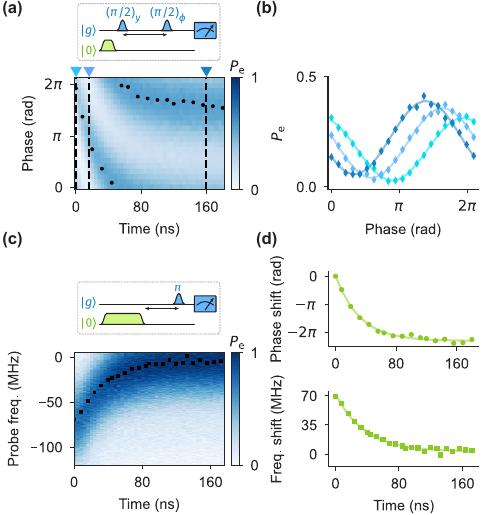}
	\caption{
    \textbf{Detecting magnon decay.}
    (a) Ramsey sequence executed during the decay of magnon population. 
    The phase of the second pulse is swept.
    Black dots: phase of the fringes obtained from sinusoidal fits.
    (b) Linecuts show phases that serve as measure of the time-dependent magnon population.
    (c) Qubit spectroscopy during magnon decay maps population onto qubit frequency.
    Black squares: resonant frequency obtained from Gaussian fits.
    (d) Time evolution of phase and frequency shifts. 
    Decay constants are extracted from exponential fits.
    }
\label{fig:fig3}
\end{figure}

Next, we turn to using the qubit as sensor for population dynamics and, as a proof-of-concept, aim to resolve the Kittel mode decay.
The underlying principle is again the dispersive shift of the qubit, which now becomes time-dependent.
We have explored two different methods by which the dynamics can be resolved by the qubit: 
First, by detecting the time-dependent phase accumulation of a superposition state;  and second, by directly measuring the time-dependent frequency shift.
To resolve the qubit's phase accumulation, we have performed a Ramsey experiment (\Cref{fig:fig3}(a),(b)).
After preparing a certain magnon population, the qubit was prepared in a superposition state, and left to evolve while the magnons decayed.
The time-dependent phase of the qubit is then a measure of the magnon relaxation.
Alternatively, we can perform time-dependent qubit spectroscopy (\cref{fig:fig3}(c)).
Here, the change in qubit resonant frequency is a direct measure of magnon relaxation.

Similarly to the case of steady-state population, we are interested in the accuracy with which magnon damping can be resolved, and how fast this information can be acquired.
To that end, we have sub-sampled the datasets presented in \cref{fig:fig3}(a) and (c) and analyzed subsets with total acquisition time of a second.
We have extracted the time-dependent phase shifts and qubit resonance frequencies, and determined their decay constants (\cref{fig:fig3}(d)).
Exponential fits yield $1/\kappa_\text{m} = \qty{34 \pm 2}{\nano\second}$ and $\qty{40 \pm 4}{\nano\second}$, in agreement with each other and with independent spectroscopic verification
(Supplement).
The fit uncertainties can be translated to a sensitivity for detecting magnon lifetime, on the order of a few $\SI{}{\nano\second\per\sqrt{\hertz}}$.
We note that the magnon lifetime is reduced by a factor of about 3 compared to typically quoted values in YIG spheres \cite{tabuchiCoherentCouplingFerromagnetic2015, xuQuantumControlSingle2023}.
We attribute this reduction to magnetic field inhomogeneity and it could be addressed by altering geometry. 

The qubit can thus be used as a sensitive detector of magnon dynamics, but it has limitations.
Here, we resolved decay for up to $\sim 650$ initialized magnons.
This limitation arises from the combination of large magnon damping rate $\kappa_\text{m}$ and dispersive shift. 
The decay of $n_\text{m}$ excitations occurs with rate $n_\text{m} \kappa_\text{m}$, producing a qubit frequency shift over a very short time. 
If qubit control is slow in comparison, the signal blurs out and dynamics at large magnon numbers are harder to detect.
The magnon number for which dynamics can be resolved could thus be increased in two ways:
First, a weaker dispersive coupling will reduce the frequency shift per time, and similarly to the case of population detection, sensitivity can be sacrificed for dynamic range.
Second, faster qubit control allows resolving shorter timescales; the timing resolution of our experiment could be improved, for example, using fast direct-synthesis solutions~\cite{stefanazziQICKQuantumInstrumentation2022a}.


\begin{figure}[t]
\centering
\includegraphics[width=3.34in]{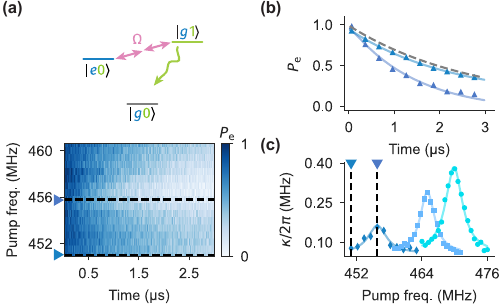}
	\caption{
    \textbf{Resonant magnon-qubit coupling.}
    (a) Driving at the half-difference frequency between transmon and magnon mode dynamically hybridizes the modes.
    The transmon becomes Purcell-limited and decays faster.
    (b) Linecuts of transmon decay at resonant (dark blue) and near-resonant (medium blue) hybridization with the magnon mode.
    Qubit decay with no drive shown dashed.
    (c) Stronger pumps result in faster qubit decay.
    The parametric pump strengths for the data shown are $\Omega_\text{qm}/2\pi = \qty{0.66\pm0.02}{\mega\hertz}$, $\qty{0.86\pm0.02}{\mega\hertz}$, and $\qty{1.11\pm0.02}{\mega\hertz}$.
    Resonances are increasingly Stark-shifted with increasing pump power~\cite{pfaff_controlled_2017}.
    }
\label{fig:fig4}
\end{figure}

So-far we have probed the magnon mode using the static dispersive $ZZ$ interaction, where qubit frequency depends on magnon occupation.
As discussed, fast magnon decay may be difficult to resolve with this coupling due to the demands on qubit control speed.
A powerful alternative approach is to utilize a tunable $XX$ coupling, where energy is exchanged resonantly.
Then, the dynamics of the system probed are mapped onto relaxation of the qubit, which can be used for detecting fast noise processes~\cite{kolkowitz_probing_2015}.
Here, we show that we can exploit the intrinsic nonlinearity of the transmon qubit to dynamically activate resonant coupling in a controlled fashion, and measure the magnon decay through the qubit relaxation. 

We have used parametric pumping to engineer an energy exchange that can be used to determine the magnon decay rate as follows.
Applying a pump to the transmon near $|\omega_\text{q} - \omega_\text{m}|/2$ realizes degenerate four-wave mixing between the two modes~\cite{mollenhauer_highefficiency_2024}.
This mixing results in energy exchange with a rate $\Omega_\text{qm}$ that is tunable by pump power.
After preparing the qubit in the excited state, we have applied a pump with fixed power for a variable time and with varying detuning from resonance (\cref{fig:fig4}(a)). 
The rapid decay of the magnon mode results in an accelerated decay rate of the qubit, $\kappa$, which can be understood as a dynamically activated Purcell effect (\cref{fig:fig4}(a) and (b)). 
When the pump is on resonance, $\kappa \equiv 1/T_1 = \Omega_\text{qm}/\kappa_\text{m}$, for $\Omega_\text{qm}/2 \ll \kappa_\text{m}$.
From a measurement of qubit decay as function of pump frequency we can independently infer $\Omega_\text{qm}$ and $\kappa_\text{m}$ for a given pump power~\cite{pfaff_controlled_2017}.
We obtain $1/\kappa_\text{m} = \qty{39 \pm 4}{\nano\second}$, in agreement with the dispersive detection method.
In addition, the steady-state population of the qubit reflects the magnon population.
In the measurements presented here, we have not applied an additional drive to the magnon mode and the qubit decays fully to the ground state, reflecting the fact that the Kittel mode excitation is close to zero in equilibrium.
We emphasize that pump strength maps magnon decay onto qubit relaxation in a controllable fashion (\cref{fig:fig4}(c)).
In this way, a wide range of different $\kappa_\text{m}$ and $\langle n_\text{m} \rangle$ could be resolved by the qubit decay.
For details on the model, see 
Supplement.


In summary, we have employed a superconducting qubit as a quantum sensor for a weakly coupled magnon mode. 
Without requiring fine-tuning of the qubit-magnon coupling, we have used quantum control and parametric pumping of a superconducting qubit to measure magnon population and decay with a high dynamic range.
Using the instrinsic dispersive coupling, we were able to count up to about 2000 magnons with few-magnon sensitivity.
Using the dispersive as well as an engineered resonant coupling we have used the qubit to measure the magnon decay.

We emphasize the simplicity of our experimental approach, which does not require intricate magnetic shielding solutions or magnetic-field compatible qubits.
While mostly inconsequential in our case, we have observed that stray magnetic fields cause magnon line broadening and a reduction in qubit coherence.
These effects could be addressed by a different magnet arrangement, for example in a Halbach geometry~\cite{halbachDesignPermanentMultipole1980}.
Tunable magnetic fields could also be accommodated in a fairly straight-forward fashion~\cite{liCoherentCouplingTwo2022}.

Our results establish superconducting qubits as high-range, high-sensitivity quantum probe for magnons.
The detection limit is set by how effectively the qubit can accumulate information about magnon properties.
Therefore, in-situ control of coupling, as we have shown here through parametric pumping, is a particularly promising approach to resolve dynamics across a wide range of populations and timescales.
Our experimental platform is compatible with the generation and stabilization of nonclassical states with high appeal for sensing, such as cat or squeezed states~\cite{leghtasConfiningStateLight2015,guoMagnonSqueezingTwotone2023}; and our methods would allow to also map phase fluctuations onto the qubit~\cite{murchReductionRadiativeDecay2013}.
Our experiment could thus be extended, for example, to resolve nontrivial decay dynamics \cite{yuan_pure_2022} or intrinsic nonlinearities.
The ability to engineer resonant interactions further offers the opportunity to realize out-of-equilibrium steady states of the qubit that could serve as sensitive probes of multi-mode systems~\cite{kitzman_phononic_2023}.
The methods employed here only rely on weak dispersive couplings between a cavity and the degree of freedom of interest, and could therefore be transferred straight-forwardly to other collective excitations.

\section*{Acknowledgments}
The research was carried out in part in the Materials Research Lab Central Facilities and the Holonyak Micro and Nanotechnology Lab at the University of Illinois. 
We thank K.~Chow and R.~Goncalves for help with fabrication, R.A.~Duine for insightful comments, and A.~Kou and J.~Lim for critical reading of the manuscript.
This work was supported by the Department of Energy (Contract No.~DE-SC0022060) and the National Science Foundation (Awards 2016136 and 2137642).

\appendix

\section{Experimental setup}

\subsection{Theoretical system description}

The Hamilton of the system (with $\hbar = 1$) is given by
\begin{equation}
\begin{aligned}
H = \omega_\text{c} \hat{c}^{\dag}\hat{c} + \omega_\text{m} \hat{m}^{\dag}\hat{m} + \omega_\text{q}\hat{q}^{\dag}\hat{q} + \frac{\alpha}{2} (\hat{q}^{\dag}+\hat{q})^2\\
+g_\text{mc}(\hat{m}^{\dag}\hat{c} + \hat{m}\hat{c}^{\dag}) + g_\text{qc}(\hat{q}^{\dag}\hat{c} + \hat{q}\hat{c}^{\dag}),
\end{aligned}
\end{equation}
where $\omega_\text{c}$, $\omega_\text{m}$, $\omega_\text{q}$ represent the frequencies and $\hat{c}$, $\hat{m}$, $\hat{q}$ represent the annihilation operators of the cavity, Kittel (magnon) mode and qubit, respectively.
The Kittel mode and the cavity are magnetic-dipole coupled with strength $g_\text{mc}$, while the qubit and the cavity are electric-dipole coupled with strength $g_\text{qc}$.
We approximate the Kittel mode as a harmonic oscillator, which is valid in the low-excitation limit~\cite{lachance-quirionHybridQuantumSystems2019}.
$\alpha$ is the anharmonicity of the transmon qubit.
We diagonalize this Hamiltonian, and in the dispersive regime it can be written as~\cite{blaisCircuitQuantumElectrodynamics2021e}
\begin{equation}
\begin{aligned}
H_\text{disp} &=  \omega_\text{c} \hat{c}^{\dag}\hat{c} + \omega_\text{m} \hat{m}^{\dag}\hat{m} + \omega_\text{q}\hat{q}^{\dag}\hat{q} + \frac{\alpha}{2} (\hat{q}^{\dag}\hat{q})^2+ 
\chi_\text{qc} \hat{q}^{\dag}\hat{q}\hat{c}^{\dag}\hat{c}\\&
+ \chi_\text{qm} \hat{q}^{\dag}\hat{q}\hat{m}^{\dag}\hat{m} +\chi_\text{mc} \hat{m}^{\dag}\hat{m}\hat{c}^{\dag}\hat{c}
\end{aligned}
\label{dispersive-hamiltonian}
\end{equation}
$\chi_{ij}$ is the dispersive frequency shift per excitation between modes $i$ and $j$.
For simplicity, the (now dressed) frequencies and normal mode operators are denoted by the same symbols as in the previous Hamiltonian.
Note that the dispersive shifts are related:
The qubit-magnon dispersive shift, $\chi_\text{qm}$, can be expressed following \cite{blaisCircuitQuantumElectrodynamics2021e} in terms of $\chi_\text{qc}$ as $\chi_\text{qm} = (g_\text{mc}/\Delta_\text{mc})^2 \chi_\text{qc}$.
Similarly, $\chi_\text{mc}$ can be expressed through $\chi_\text{qm}$.

\subsection{Device design}
\label{app:hqs-design}

\begin{figure}[t]
    \centering
    \includegraphics[width=3.34in]{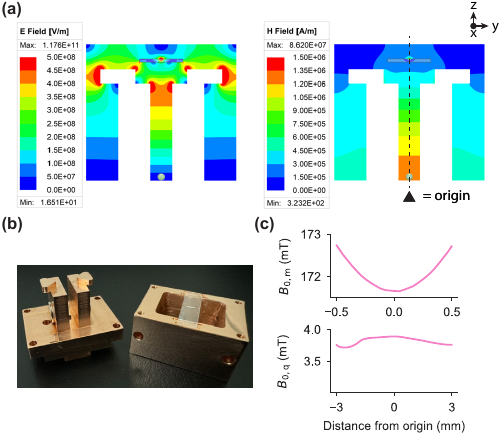} 
    \caption{
    \textbf{Device design.}
    (a) Finite element simulations of electric ($E$) and magnetic ($H$) field for the cavity mode at center of the cavity, in the YZ plane.
    Oscillating $E$ fields are concentrated at the qubit position on top; 
    oscillating $H$ fields are maximal near the YIG sphere position.
    Dashed line is used to refer to the origin (center) of the cavity.
    (b) Photograph of the main components of the device.
    The cavity is made of three Cu pieces that are stacked on top of each other and bolted together with screws.
    Bottom piece is shown left, middle piece (incl qubit chip) on the right.
    A lid that mounts to the top of the middle piece is not shown.
    YIG sphere barely visible at the center of the bottom piece.
    One of the permanent magnets used is visible on the side of the bottom piece.
    (c) Simulated static magnetic field ($B_0$) from the permanent magnets at position of the YIG sphere (top panel) and qubit chip (bottom panel) on a line along the y-axis.
    }
    \label{fig:figs1}
\end{figure}

Finite element simulations (obtained from Ansys HFSS) of the lumped-element cavity are shown in \Cref{fig:figs1}(a). 
The spatially separated cavity field components enable coupling to qubit and magnon at different locations, minimizing the effect of stray field from the permanent magnet on the qubit. 
The qubit-cavity coupling strength and the corresponding dispersive shift are evaluated by calculating the energy participation ratio of the transmon junction~\cite{minevEnergyparticipationQuantizationJosephson2021a}. 
The magnon-cavity coupling is evaluated with the magnetic field value integrated over the YIG sphere volume normalized to half-photon energy~\cite{angerer_collective_2016a}.  
A picture of the physical device is shown in~\cref{fig:figs1}(b).

We use small disc-shaped neodymium magnets to provide a near-uniform static magnetic field over the \qty{1}{\milli\meter} diameter YIG sphere volume.
From a finite element simulation of the magnets (using Ansys Maxwell) in the cavity we estimate a \qty{172}{\milli\tesla} magnetic field at the sphere location, resulting in a prediction of the Kittel mode frequency of \qty{4.82}{\giga\hertz}, while maintaining a magnetic field of less than \qty{4}{\milli\tesla} at the qubit's location, as shown in \cref{fig:figs1}(c).
The permanent magnets near the YIG sphere introduce a slight inhomogeneity in the magnetic field across the \qty{1}{\milli\meter} range, with a simulated center-to-edge inhomogeneity of about \qty{1}{\milli\tesla} (\cref{fig:figs1}(c)).

\begin{figure}[t]
\centering
\includegraphics[width=3.34in]{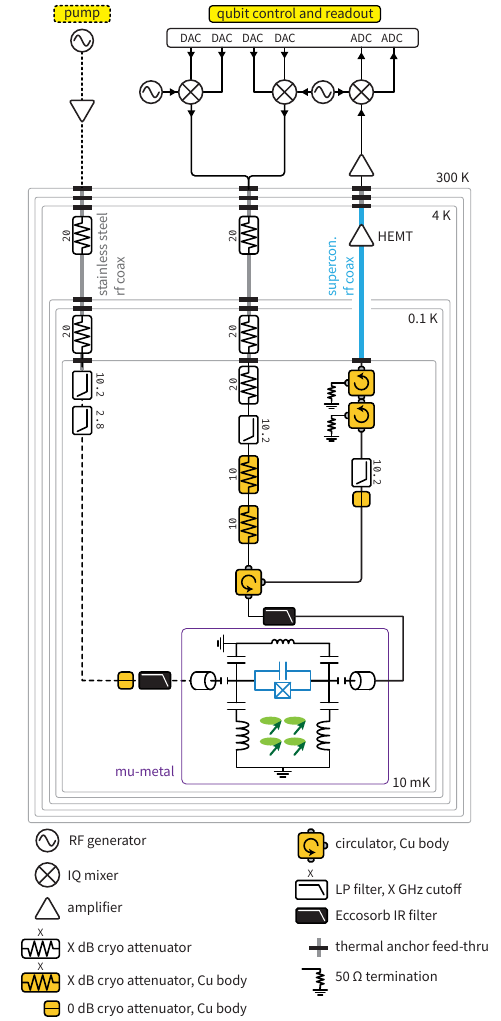} 
	\caption{
    \textbf{Cryogenic setup.}
    Qubit, cavity, and magnon control signals were synthesized using a commercial quantum control solution (Quantum Machines OPX) and up/down-converted using external local oscillators (Signalcore) and IQ mixing.
    Pumps were supplied using a gated microwave generator (Rohde \& Schwarz SGS100A).
    }
\label{fig:figs2}
\end{figure}
\begin{figure}[t]
\centering
    \includegraphics[width=3.35in]{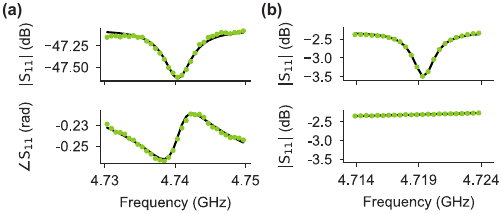}
	\caption{
    \textbf{Spectroscopic magnon mode characterization.}
    (a) $S_{11}$ measurement of the Kittel mode at $T = \qty{10}{\milli\kelvin}$.
    (b) RT $S_{11}$ measurement of the cavity near Kittel mode frequency with (top plot) and without (bottom plot) magnet.
    Differences in frequency and linewidth of the mode are due to (1) magnet alignment shifts in the cooldown/warmup process of the device, and (2) the absence of a superconducting qubit in the RT measurements; the chip was removed for the RT measurements.
    The linewidth reported in \Cref{table:table1} is from a fit to a reflection model to the cryogenic data.
    }
\label{fig:figs3}
\end{figure}

\subsection{Cryogenic setup}
\label{app:cryo-setup}
The setup is assembled and cooled down to $\qty{10}{\milli\kelvin}$ in an \textit{Oxford Instruments Triton 500} dilution refrigerator.
A schematic of the measurement setup is shown in \cref{fig:figs2}.

\subsection{Device characterization}
Device parameters are summarized in \cref{table:table1}.
Qubit and cavity characterization was performed with standard spectroscopic and pulsed relaxation/dephasing measurements.

\begin{table}[h!]
    \centering
    \begin{tabular}{>{\raggedright\arraybackslash}p{5cm} c c}
        \hline
        \hline
        Qubit ($\hat{q}$) & Symbol & Value\\
        \hline
        Mode frequency   &  $\omega_\text{q}/2\pi$   & $\qty{3.87}{\giga\hertz}$   \\
        Relaxation time   &  $T_1$   & $\qty{2.78 \pm 0.07}{\micro\s}$   \\
        Ramsey decay   &  $T_\text{2R}$   & $\qty{4 \pm 1}{\micro\s}$   \\
        Hahn echo decay &  $T_\text{2E}$ & $\qty{5 \pm 2}{\micro\s}$\\
        \hline
        Cavity ($\hat{c}$) &  &  \\
        \hline
        Mode frequency   &  $\omega_\text{c}/2\pi$   & $\qty{4.56}{\giga\hertz}$   \\
        \hline
        Magnon ($\hat{m}$) &  &  \\
        \hline
        Mode frequency   &  $\omega_\text{m}/2\pi$   & $\qty{4.74}{\giga\hertz}$   \\
        Mode linewidth & $\kappa_\text{m}/2\pi$ &  $\qty{4.81}{\mega\hertz}$\\
        \hline
        Cross-Kerr ($\chi_{ij}$) & &  \\
        \hline
        Qubit and cavity   &  $\chi_\text{qc}/2\pi$   & [\qty{1}{\mega\hertz}]  \\
        Qubit and magnon & $\chi_\text{qm}/2\pi$ &  \qty{67\pm 1}{\kilo\hertz}\\
    \end{tabular}
    \caption{\textbf{System parameters.}
    Value in square brackets is obtained from simulation.
    All other quantities are obtained from measurement.
    }
    \label{table:table1}
\end{table}

Kittel mode frequency and linewidth were measured directly using a vector network analyzer (VNA) (\cref{fig:figs3}).
While the hybridization (and thus signal) is weak, the magnon mode can still be resolved in spectroscopy.
We note that this characterization is not required by our experiment, but serves as an independent verification of the Kittel mode.
To confirm that the observed mode is indeed the Kittel mode, we have also performed the VNA measurement at room temperature (RT), with and without magnets.
Disappearance of the signal without magnets is strong proof that the observed mode is the Kittel mode.

\section{Magnon counting sensitivity}
\label{app:counting}

\subsection{Magnon number calibration}
\label{app:number-calibration}
To calibrate the magnon number in our setup, we use two measurements as discussed in the main text.
We determine $\chi_\text{qm}$ and $\langle n_\text{m} \rangle$ using Stark shift and dephasing of the qubit.
The Stark shift is  

\begin{equation}
    \Delta f_{\text{q}} = \chi_{\text{qm}} \langle n_\text{m} \rangle.
\end{equation}

For the qubit dephasing, we can write the total dephasing rate as~\cite{gambettaQubitphotonInteractionsCavity2006}

\begin{equation}
    \Gamma_\text{q} = \gamma_\text{2}^0 + 2 \langle n_\text{m} \rangle   \kappa_\text{m}\frac{\chi_\text{qm}^2}{(\kappa_\text{m}^2 + \chi_\text{qm}^2)}.
\end{equation}
Here, $\gamma_\text{2}^0$ is the `bare' transmon dephasing rate with zero magnon population, and can be measured independently.
Note that the magnon occupation number $\langle n_\text{m} \rangle$ is proportional to the pump power, $\langle n_\text{m} \rangle = c_{\text{pump}} P$, where $c_{\text{pump}}$ is determined by the attenuation in the fridge wiring. 
There are only two unknowns ($c_{\text{pump}}$ and $\chi_\text{qm}$); these equations thus allow us to extract the value of $\chi_{\text{qm}}$ and calibrate the magnon number by measuring the Stark shift and qubit dephasing rate as a function of pumping power. 


\subsection{Experimentally obtained sensitivity}
\label{app:sensitivity-expt}

To evaluate the sensitivity $S(n_\text{m})$  of our setup as described in the main text, we express SNR based on signal and noise (Fig.~2(c) in the main text) as:
\begin{equation}
    \text{SNR} =     \frac{|P_\text{e} - P_\text{e}^{'} |}{\sqrt{\sigma_{P_\text{e}}^{2} + \sigma_{P_\text{e}^{'}}^{2}}}
    \label{snr equation1}
\end{equation}
Here, $\sigma_{P_\text{e}}$, $\sigma_{P_\text{e}^{'}}$ are the standard deviations of the mean when evaluating $P_\text{e}$ and $P_\text{e}^{'}$, with uncertainty arising from noisy qubit readout.
Theses values are related to the standard deviation defined in the main text as $\sigma_{P_\text{e}} = \sigma/\sqrt{N}$, where N is the number of samples used to evaluate $P_\text{e}$.
To determine \textit{S} as a function of magnon population, it is necessary to evaluate the signal and noise for arbitrary $n_\text{m}$ and $n_\text{m}^{'}$.

The signal is obtained from qubit spectroscopy as follows.
We use a Gaussian distribution to model the qubit response $P_{\text{e}}(n, n_{\text{m}})$~\cite{gambettaQubitphotonInteractionsCavity2006} as
\begin{equation}
    P_{\text{e}}(n, n_{\text{m}}) = 
        P^{n_{\text{m}}}_{\text{e}} 
            \exp \left(\frac{(n - n_{\text{m}})^2}{2 \Sigma_{n_\text{m}}^{2}}\right),
\end{equation}
where $P^{n_{\text{m}}}_{\text{e}}$ is the qubit excited-state probability after applying a $\pi$ - pulse at frequency $\omega_\text{q} + n_\text{m}\chi_\text{qm}$.
$\Sigma_{n_\text{m}}$ is the standard deviation of a Gaussian fit of the qubit response versus probe frequency, after preparing $n_\text{m}$ magnons.
Note that the SNR can then be rewritten as
\begin{equation}
    \text{SNR} = \frac{|P_{\text{e}}(n_{\text{m}}, n_{\text{m}}) - P_{\text{e}}(n_{\text{m}} + S(n_{\text{m}}), n_{\text{m}}) |}{\sqrt{\sigma_{P_\text{e}}^{2} + \sigma_{P_\text{e}^{'}}^{2}}}.
    \label{snr equation2}
\end{equation}

\begin{figure}[t]
\centering
    \includegraphics[width=3.35in]{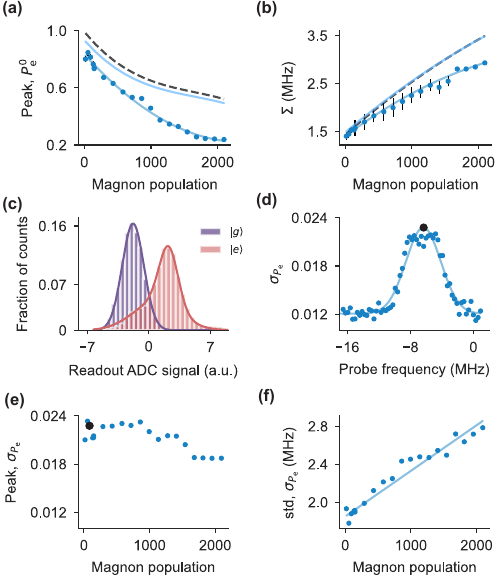}
	\caption{
    \textbf{Sensitivity extraction.}
    (a), (b) $P_\text{e}^{0}$ and $\Sigma$ as function of magnon number.
    Dark blue dots: parameters from Gaussian fits to spectroscopy curves shown in Fig.~2 of the main text.
    Dark blue solid line: polynomial fit to data points.
    Light blue line:  simulation for qubit with measured coherence times.
    Gray dashed line: simulation with ideal qubit.
    (c) Histogram: Qubit readout histogram with qubit prepared in ground (purple) and excited state (orange).
    Solid lines: Gaussian (purple) and double-Gaussian (orange) fits of ground and excited state readout, respectively. 
    Double-Gaussian fit is required because the qubit decays to the ground state during readout.
    (d) Standard error of the mean for $P{\text{e}}$ w.r.t.\ probe frequency. 
    Solid line: Gaussian fit.
    Black point: peak value.
    (e) Extracted peak of the Gaussian fits described in (d).
    Black point corresponds to the peak value obtained for (d).
    (f) extracted standard deviation of $\sigma_{P_\text{e}}$ from the Gaussian fit.
    Solid line: polynomial fit.
    }
\label{fig:figs4}
\end{figure}

From the Stark shift measurements (Fig.~2(a) in the main text), we extract the peak value ($P^{n_{\text{m}}}_{\text{e}}$) and standard deviation ($\Sigma_{n_{\text{m}}}$) from a Gaussian fit at each measured magnon number (\cref{fig:figs4}(a),(b)).
By interpolating both as a function of magnon population using a second-order polynomial, we obtain parameters for a Gaussian for arbitrary magnon number, allowing us to evaluate the signal for arbitrary $n_\text{m}$.

To estimate the noise for arbitrary $n_\text{m}$, we evaluate the standard deviation of the mean qubit excited state probability, ($\sigma_{P_\text{e}}$), versus probe frequency for each measured magnon occupation; for reference, see \Cref{fig:figs4}(c) for readout distributions after preparing ground and excited states of the qubit.
$\sigma_{P_\text{e}}$ versus probe frequency for one particular magnon occupation is shown in \Cref{fig:figs4}(d).
To extract the noise profile for arbitrary magnon number using interpolation, we fit this profile with an (empirical) Gaussian.
The extracted peak of $\sigma_{P_\text{e}}$ and standard deviation of $\sigma_{P_\text{e}}$ from this fit are plotted in \cref{fig:figs4}(e),(f) as a function of magnon population.
It is worth noting that the noise estimate does not significantly impact the sensitivity values since the noise levels for the ground and excited states are comparable.
Thus, we estimate noise by the average peak value and a linear fit of the standard deviation to generating Gaussian $\sigma_{P_\text{e}}$ versus probe frequency for arbitrary $n_\text{m}$.

We follow the usual definition of sensitivity from \cite{degenQuantumSensing2017}, which states that it is the minimum detectable signal that yields unit SNR in \qty{1}{\s} total measurement time.
In our case, the total measurement time for the magnon detection sequence is $T  = \text{N}\tau \sim \qty{32}{\milli\s}$ (with $\tau\sim\qty{32}{\micro\s}$, N = 1000).
Hence, using the definition of \cref{snr equation2} and making use of $\mathrm{SNR} \propto \sqrt{\text{N}}$, we compute sensitivity by finding the value $S(n_\text{m}) \equiv n_\text{m}^{'}-n_\text{m}$ for which $\mathrm{SNR} = 0.18$ with $T =\qty{32}{\milli\s}$. 
This procedure results in the sensitivity values as a function of magnon population presented in the main text.

\subsection{Theoretical limits of sensitivity}
\label{app:counting-thy}

To determine the theoretical limits on sensitivity, we perform numerical simulations of the system in the time-domain, accounting  for measured coherence times and magnon number dependent dephasing rates of the qubit following \cite{gambettaQubitphotonInteractionsCavity2006}.
We match the readout to the experimentally obtained parameters (\cref{fig:figs4}(c)).
This provides simulated values for both standard deviation ($\Sigma$) and peak ($P^{n_{\text{m}}}_{\text{e}}$) as a function of magnon population (\cref{fig:figs4}(a),(b)).

In addition, we simulate an ideal qubit, isolating the effects of decoherence on sensitivity, while preserving comparable readout conditions. 
Unlike the experimental scenario, the ideal qubit's infinite coherence prevents any population leakage from the excited state to the ground state during the measurement, thereby modifying the readout noise. 
We fit the measured ground-state histogram with a Gaussian distribution and the excited-state histogram with a double-Gaussian distribution (\cref{fig:figs4}(c)), to extract the standard deviations of a pure ground/excited state measurement. 
These signal and noise values are then incorporated into the signal-to-noise ratio (SNR) definition, yielding the sensitivity curves presented in Fig.~2(e) of the main text.

\section{Parametrically activated resonant coupling}
\label{app:parametric-expt}

\begin{figure}[t]
\centering
    \includegraphics[width=3.35in]{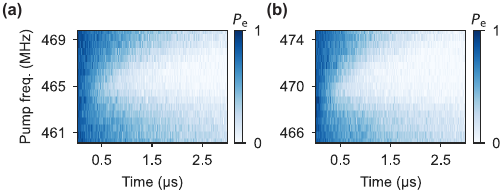}
	\caption{
    \textbf{Resonant magnon-qubit coupling: additional data.}
    (a) and (b) are the raw data from which the medium blue and dark curves in Fig.~4(c) of the main text were extracted.
    }
\label{fig:tuneup-parametric-swap}
\end{figure}

The transmon can be used as a four-wave mixer due to its fourth order nonlinearity.
A pump at an appropriate frequency enables a parametric mode conversion interaction with a controllable coupling strength $\Omega_\text{qm}/2$, described by the Hamiltonian
\begin{equation}
    H_\text{int} = \frac{\Omega_\text{qm}}{2} \hat{q}^\dag \hat{m} + \text{h.c.}.
\end{equation}
Particularly in our case, the pump frequency is given by $\omega_\text{pump} = |\omega_\text{q} - \omega_\text{m}|/2$.
See, for instance, \cite{mollenhauer_highefficiency_2024} for details on the derivation.
The strength of this conversion interaction is set by the pump power.

To model the accelerated decay rate of the qubit we follow \cite{pfaff_controlled_2017} and find the time dependence of $\hat{q}$,

\begin{equation}
\begin{aligned}
    \hat{q}(t) &= \frac{\hat{q}(0)}{\beta} e^{-\frac{\gamma t}{4}} \left(\beta \cosh{\left(\frac{t\beta}{4}\right)} + \gamma  \sinh{\left(\frac{t\beta}{4}\right)}\right)
\end{aligned}
\end{equation}
where
\begin{subequations}
    \begin{align}
        \gamma &= \kappa_\text{m} + 2 \text{i} \delta\\
        \beta &= \sqrt{\gamma^2 - (2\Omega_\text{qm})^2},
    \end{align}
\end{subequations}
and $\delta$ is a small detuning from the resonance condition of the conversion process.
\\
To understand the dynamically activated Purcell effect we consider the case $\delta = 0$ and $\Omega_\text{qm}/2 \ll \kappa_\text{m}$, yielding the approximate solution
\begin{equation}
    \begin{aligned}
        \hat{q} \approx \hat{q}(0) e^{-\Omega_\text{qm}^2t/2\kappa_\text{m}}.
    \end{aligned}
\end{equation}
Clearly, the decay rate of qubit in this case is $\kappa = \Omega_\text{qm}^2/\kappa_\text{m}$.
For $\delta \neq 0$, we can write evolution of $\hat{q}$ as:
\begin{equation}
    \hat{q}\approx\hat{q}(0)e^{-\frac{t}{2}\frac{\Omega_\text{qm}^2}{\gamma}} \equiv \hat{q}(0) e^{-\frac{t}{2}\gamma_\text{conv}} 
\end{equation}
with complex $\gamma_\text{conv}$,
\begin{equation}
    \gamma_\text{conv} = \frac{\Omega_\text{qm}^2(\kappa_\text{m} - 2 \text{i} \delta)}{\kappa_\text{m}^2+(2\delta)^2}.
\end{equation}
The real part of $\gamma_\text{conv}$ is the qubit decay rate, $\kappa$.
$\kappa$ is a Lorentzian with FWHM $\kappa_\text{m}$, and centered around $\delta=0$.
The data in Fig.~4(c) of the main text are fit to this function.
The medium blue and light blue Lorentzian fits shown in main text Fig.~4(c) 
are derived from the data shown in \cref{fig:tuneup-parametric-swap}.

\end{document}